\documentclass[preprint,aps]{revtex4}
\usepackage{graphicx}
\begin{document}

\title{Novel steady state of a microtubule assembly in a confined geometry}

\author{Bindu S. Govindan}
\affiliation{Applied Biosciences Center,
Virginia Polytechnic Institute  and State University,
Blacksburg, VA 24061-0356, U. S. A.}

\author{William B. Spillman, Jr.}

\affiliation{Applied Biosciences Center and Department of Physics,
Virginia Polytechnic Institute  and State University,
Blacksburg, VA 24061, U. S. A.
}

\date{\today}
\begin{abstract}

We study the steady state of an assembly of microtubules in a
confined volume, analogous to the situation inside a cell where the cell
boundary forms a natural barrier to growth. We show that the dynamical
equations for growing and shrinking microtubules predict the existence
of two steady states, with either exponentially decaying or exponentially
increasing distribution of microtubule lengths. We identify the regimes in
parameter space corresponding to these steady states. In the latter case, the 
apparent 
catastrophe frequency near the boundary was found to be significantly larger than 
that in the interior. Both the exponential distribution of lengths and the
increase in the catastrophe frequency near the cell margin is in excellent
agreement with recent experimental observations.
\end{abstract}

\maketitle

Microtubules are long, rigid polymers which play an important role in
several cellular processes. They are highly dynamic structures, which
are perpetually in a state of growth or shrinkage, and switch
stochastically between these states. This behavior is called dynamic
instability\cite{ref3,ref4}. The basic monomer
unit of a microtubule is a dimer of $\alpha$ and $\beta$ tubulin, which is
approximately 8 nm in length. The
$\alpha-\beta$ dimers are arranged head-to-tail along a microtubule in
protofilaments (usually 13 per microtubule). 

The highly dynamic nature of microtubules originates from the
hydrolysis of $\beta$-tubulin bound GTP. Following hydrolysis, the GTP
is converted to GDP \cite{ref2}, and the GDP-bound
tubulin does not polymerize as well as its GTP counterpart. 
When the advancing hydrolysis front reaches the
growing end of a microtubule, it starts de-polymerizing and the protofilaments
start peeling off, releasing the GDP-tubulin into solution. Outside a
microtubule, reverse hydrolysis takes place and polymerization events
start all over again.  Thus the microtubule constantly switches
between phases of growth and shrinkage.
The stochastic transition from growth to shrinkage is 
called {\it catastrophe} and the reverse transition is called {\it rescue}. 
The relative rates of catastrophe and rescue, combined with
the velocities of growth and shrinkage determine the character of a
given population of microtubules\cite{ref5,ref6,ref7}.

The first theoretical model of microtubule dynamics based on the
dynamical instability mechanism was developed by Dogterom and
Leibler\cite{DOGTEROM}. In this model, microtubules are assumed to nucleate and grow
from a flat substrate, and the dynamics is characterized by the
velocity of growth ($v_{g}$) and shrinkage ($v_{s}$), and the
frequencies of catastrophe ($\nu_{c}$) and rescue ($\nu_{R}$). In the
absence of any boundary which restricts the growth, a steady state was
achieved when $\nu_R p_g < \nu_C p_s$, characterized by an
exponentially decaying distribution of lengths. When this condition
was not satisfied, no steady state was reached, and the length
distribution was Gaussian, with mean length increasing linearly with
time, and the width evolving diffusively.

Inside cells, the microtubule growth is constrained by the presence of
the cell boundary. Experimental observations have shown that the
parameters of dynamics show a strong dependence on the proximity to
cell boundary\cite{VOROB}. In particular, the catastrophe frequency was markedly
higher near the periphery, compared to the cell interior. The obvious
explanation for this difference was that the growing microtubule loses
its `GTP cap' upon hitting the cell boundary and is transformed to a
shrinking state. In addition, the length distribution of microtubules
was found to be exponentially increasing, with a possible dip near the
boundary.

In this brief communication, we show that the exponentially increasing length
distribution of microtubules can be understood from the
Dogterom-Leibler equations, and is a new steady state which is a
direct consequence of the presence of the cell boundary. We compute
the steady state distribution exactly, and find excellent agreement
with experimental observations. We also show that the observed increase in the
apparent catastrophe frequency near the cell margin can be understood quantitatively
within this model.

Let us consider a set of mcrotubules nucleating from a substrate, and growing 
by the addition of tubulin dimers in the direction perpendicular to the plane of the 
substrate (the $z$ axis). For simplicity, we ignore the three-dimensional structure 
of individual microtubules, and treat them as one-dimensional polymers. Nucleation is assumed 
to take place at empty nucleation sites at a rate $\nu$. A microtubule in growing state adds 
T-tbulin at a rate $p_{g}$ per unit time, and a microtubule in shrinking state loses tubulin 
at a rate $p_s$ per unit time. Also, a microtubule in growing state switches to shrinking 
state at a rate $\nu_{c}$ (catastrophe frequency), and a microtubule in shrinking state 
switches to growing state at a rate $\nu_{R}$ (rescue frequency). Both rescue and catastrophe 
are assumed to be purely stochastic events. 

Our principal aim in this paper is to study explicitly the steady state of
the system in the presence of a boundary. We assume that this boundary is located at 
$z=l^{*}$.  We denote by $p_{+}(l,t)$ the fraction of sites in the substrate which has 
a microtubule of
length $l$ at time $t$ in growing state, and $p_{-}(l,t)$ denote the same fraction in shrinking
state. By convention, the fraction of vacant sites in the lattice at time $t$ is denoted
$p_{-}(0,t)$, and $p_{+}(0,t)=0$ at all times $t$. The discrete equations for the dynamics of
this assembly, including growth, shrinkage, catastrophe and rescue events are given by
 
\begin{equation}
\frac{\partial p_{-}(0,t)}{\partial t}=-\nu
p_{-}(0,t)+p_{s}p_{-}(1,t)~~~~~; l=0
\label{eq:PML0}
\end{equation}

and

\[
p_{+}(0,t)=0
\]

\begin{equation}
\frac{\partial p_{+}(1,t)}{\partial t}=\nu
p_{-}(0,t)-p_{g}p_{+}(1,t)-\nu_C p_{+}(1,t)\nu_{R}p_{-}(1,t)~~~~; l=1
\label{eq:PPL1}
\end{equation}

\begin{equation}
\frac{\partial p_{+}(l,t)}{\partial t}=p_{g}[
p_{+}(l-1,t)-p_{+}(l,t)+\nu_{R} p_{-}(l,t)-\nu_{c}p_{+}(l,t)~~~~;
1<l<l^{*}
\label{eq:PPL}
\end{equation}

\begin{equation}
\frac{\partial p_{-}(l,t)}{\partial t}=p_{s}[
p_{+}(l+1,t)-p_{-}(l,t)+\nu_{c} p_{+}(l,t)-\nu_{R}p_{-}(l,t)~~~~;1\leq l<l^{*}
\label{eq:PML}
\end{equation}

The presence of the boundary affects the dynamics of the system in the following way: When
a growing microtubule reaches a length $l^{*}$, it is instantaneously transformed to the
shrinking state with length $l^{*}$. The equations representing this process are give by,

\begin{equation}
\frac{\partial p_{-}(l^{*},t)}{\partial t}=p_{g}
p_{+}(l^{*}-1,t)-p_{s}p_{-}(l^{*},t)
\label{eq:PML*}
\end{equation}

\[
p_{+}(l^{*},t)=0
\]

To find the steady state of the system, let us put all time derivatives to zero. 
From Eq.\ref{eq:PML0} and Eq.\ref{eq:PPL1} we get the follwing relations.

\begin{equation}
p_{-}(0)=\frac{p_s}{\nu} p_{-}(1)
\label{eq:REL1}
\end{equation}

\begin{equation}
p_{+}(1)=\frac{\nu p_{-}(0)+\nu_{R} p_{-}(1)}{p_{g}+\nu_{c}}
\label{eq:REL2}
\end{equation}

After combining Eq.\ref{eq:REL1} and Eq.\ref{eq:REL2}, we find that

\begin{equation}
p_{-}(1)= \frac{p_g+\nu_{c}}{p_{s}+\nu_{R}}p_{+}(1),
\label{eq:REL12}
\end{equation}

and, after using Eq.\ref{eq:REL1} again, 

\begin{equation}
p_{-}(0)=\frac{p_s}{\nu}\frac{p_{g}+\nu_{c}}{p_s+\nu_{R}} p_{+}(1)
\label{eq:REL3}
\end{equation}

For $l >1$, we find the following relation between $p_{+}(l)$ and $p_{-}(l)$ 
from Eq.\ref{eq:PPL} and Eq.\ref{eq:PML} (using only $l>1$ in Eq.\ref{eq:PML}).

\[
p_{-}(l+1)-p_{-}(l)=\frac{p_g}{p_{s}}[p_{+}(l)-p_{+}(l-1)] ~~~~~~l\ge 2
\label{eq:DERIV}
\]

The general solution of this equation is

\begin{equation}
p_{-}(l)=\frac{p_g}{p_s}p_{+}(l-1)+C ~~~~ l\ge 2
\label{eq:PMPP}
\end{equation}

where $C$ is an unknown constant. After substituting Eq.\ref{eq:PMPP} in Eq.\ref{eq:DERIV}, 
we obtain the following equation for $p_{+}(l)$.

\begin{equation}
p_{g}[p_{+}(l-1)-p_{+}(l)+\nu_{R}[\frac{p_{g}}{p_{s}}p_{+}(l-1)+]C=\nu_{c}p_{+}(l)~~~~l\ge 2
\label{eq:PP}
\end{equation}

A trial solution to this equation has the form 

\begin{equation}
p_{+}(l)=Aa^{l}+B~~~~; l\ge 1
\label{eq:SOLUTION}
\end{equation}

We now substitute this solution into Eq.\ref{eq:PP}, and after equating terms with the
same power of $l$, we obtain the following expressions for the constants $a$ and $B$.

\begin{eqnarray}
a=\frac{1+\frac{\nu_R}{p_s}}{1+\frac{\nu_{c}}{p_g}}\\
\label{eq:EQA}
B=\frac{\nu_{R}p_{s}}{\nu_{c}p_{s}-\nu_{R}p_{g}}C
\label{eq:EQB}
\end{eqnarray}

The constant $C$ may now be determined as follows: From Eq.\ref{eq:REL3}, for $l=1$, we have

\begin{equation}
p_{s}[p_{-}(2)-p_{-}(1)=\nu_{R}p_{-}(1)+\nu_Cp_{+}(1),
\label{eq:CRUCIAL}
\end{equation}

whereas from Eq.\ref{eq:PMPP} we have another relation:

\begin{equation}
p_{-}(2)=\frac{p_g}{p_s}p_{+}(1)+C
\label{eq:C11}
\end{equation}

We now substitute Eq.\ref{eq:REL12} and Eq.\ref{eq:C11} into Eq.\ref{eq:CRUCIAL} and
solve for $C$, which gives $C=0$. From Eq.\ref{eq:EQA}, this also implies $B=0$. It remains 
to determine the constant $A$, which is found using normalization:

\begin{equation}
\Sigma_{l=0}^{l^{*}}p_{-}(l)+\Sigma_{l=1}^{l^{*}-1}p_{+}(l)=1
\label{eq:NORM}
\end{equation}

which may be written as 

\begin{equation}
p_{-}(0)+p_{-}(1)+(1+\frac{p_g}{p_s})\Sigma_{l=1}^{l^{*}-1}p_{+}(l)=1
\label{eq:NORM1}
\end{equation}

after using Eq.\ref{eq:PMPP}. We now use Eq.\ref{eq:REL12}, Eq.\ref{eq:REL3} and 
Eq.\ref{eq:SOLUTION} in Eq.\ref{eq:NORM1}. The final result is

\begin{equation}
A=\left[\frac{p_g}{\nu}+\frac{p_g}{p_{s}+(\frac{p_g+p_s}{p_s})[\frac{a^{l^*}-a}{a-1}]}\right]^{-1}
\label{eq:AA}
\end{equation}

The solution in Eq.\ref{eq:SOLUTION} can also be written as

\begin{equation}
p_{+}(l)=A e^{\alpha l}~~~; \alpha=log\left[\frac{1+\frac{\nu_R}{p_s}}{1+\frac{\nu_C}{p_g}}\right]
\label{eq:SOLUTION1}
\end{equation}

The complete length distribution may now be written explicitly:

\begin{eqnarray}
p(l=1)=A(\frac{p_g}{p_s}+a)\\
p(l)=A(1+\frac{p_g}{ap_s})e^{\alpha l}~~~~;1 < l < l^{*}\nonumber\\
p(l=l^{*})=\frac{p_g}{p_s}\frac{A}{a}e^{\alpha l^{*}}
\label{eq:COMPLETE SOLN}
\end{eqnarray}

If $\frac{\nu_R}{p_{s}} < \frac{\nu_C}{p_g}, \alpha <0$ and we have an exponentially 
decaying solution. On the other hand, if $\frac{\nu_R}{p_{s}} > \frac{\nu_C}{p_g}, a>1$ and 
$\alpha >0$, and we have an exponentially increasing steady state distribution of lengths. 

It is interesting to look at the behavior of the solution in the limit $l^{*}\to\infty$.
From Eq.\ref{eq:AA}, we see that, in this limit, a steady state is possible only if $a <1$.
For, if $a>1$, then $A\sim a^{-l^{*}}$ for large $l^{*}$ and vanishes as $l^{*}\to\infty$. 
For $a<1$ and $l^* \to \infty$, $\frac{a^{l^{*}}-a}{a-1} \to \frac{a}{1-a}$, and so

\begin{equation}
A_{l^{*}\to\infty}=[\frac{p_g}{\nu}+\frac{p_g}{p_{s}+(\frac{p_g+p_s}{p_s})\frac{a}{1-a}}]^{-1}~~~;a < 1
\label{eq:AINF}
\end{equation}

The exponentially decaying steady state length distribution when $l^{*}=\infty$,
with  $\frac{\nu_R}{p_{s}} < \frac{\nu_C}{p_g}$, has
been predicted by Dogterom and Leibler in an earlier work\cite{DOGTEROM}. The novel 
feature in the finite $l^{*}$
case is the steady state with exponentially increasing distribution of lengths when 
$\frac{\nu_R}{p_{s}} > \frac{\nu_C}{p_g}$.

The exponentially increasing distribution of microtubule lengths has indeed been observed
in experiments with real cells. Direct observation of microtubules inside cells has been
made possible recently\cite{VOROB}. These experiments, done on centrosome-containing cytoplasts, observed 
almost persistent growth of microtubules almost up to the cell boundary. However, the catastrophe
rate showed a dramatic increase within 
a zone about 3$\mu$m near the cell margin (0.08$s^{-1}$, compared
to 0.005 $s^{-1}$ in the cell interior). The other parameters describing the microtubule dynamics
were $\nu_{R}\approx 0.12s^{-1}, v_{g}\approx 17.8 \pm 13.8\mu m /min, v_{s}\approx 28.8 \pm 14.1\mu m/min$. The parameters
$p_{g}$ and $p_{s}$ are related to $v_{g}$ and $v_{s}$ as $p_{g}=\frac{v_g}{\delta}$ and
$p_{s}=\frac{v_s}{\delta}$, where $\delta$ is the unit of length for our
effective one-dimensional polymers. Since a microtubule consists of 13 protofilaments, and
the length of a single tubulin dimer is 8 nm, this length scale is 
$\delta= 8 nm/13\approx 0.6 nm$.

As a first test of our model, we compute the increase in catastrophe frequency near
the cell margin. Let us consider all microtubules with length between $l_{1}$ and $l^*$. The total
number of such microtubules is given by $N=\sum_{l_1}^{l^*}p(l)$. Using the expression for $p(l)$
from Eq.\ref{eq:COMPLETE SOLN}, we find that 

\[
N=\frac{A}{\alpha}[1+\frac{p_g}{ap_s}][e^{\alpha l^*}-e^{\alpha l_1}]
\]

The number of microtubules in this set undergoing catastrophe per unit time is given by

\[
N^{*}=\nu_{c}N+p_{g}p_{+}(l^{*}-1),
\]

where the first term is the standard catastrophe term, and the second term represents the
additional catastrophe events arising from the microtubules hitting the boundary.
The apparent catastrophe frequency is given by $\nu_{c}^{*}=\frac{N^*}{N}$. After substituting
for $p(l)$ and $p_{+}(l^{*}-1)$, we find that

\begin{equation}
\nu_{c}^{*}=\nu_{c}+\frac{p_{g}\alpha}{1+\frac{p_g}{ap_{s}}}\frac{e^{-\alpha}}{1-e^{-\alpha\Delta}}
\label{eq:APPCAT}
\end{equation}

where $\Delta=\frac{l^{*}-l_{1}}{\delta}$. After substituting for all the numerical 
values and for $l^{*}-l_{1}\simeq 3\mu m$ as in experiments, we find
that $\nu_{c}^{*}\simeq 0.0964 s^{-1}$. This is in excellent agreement with the 
experimentally measured value of 0.08 $s^{-1}$.

It is also interesting to compare the experimentally measured value of $\alpha$ with
the theoretical value. The observed steady state length distribution was found to
fit well with an exponential function $P(l)\sim e^{\gamma l}$ with $\gamma^{-1}\simeq 
5.8 \mu m$\cite{VOROB}. We can convert this value to dimensionless units by 
multiplying with our
unit of length, which gives $\alpha_{exp}=\delta\gamma^{-1}\simeq 1.03\times 10^{-4}$.
The theoretical value is found from Eq.\ref{eq:SOLUTION1}, using the measured values of
all the parameters, and turns out to be $\alpha\approx 1.5\times 10^{-4}$. This is also in
in very good agreement with the 
experimental value. The discrepancy between the computed and observed values may be 
attributed 
to the significant experimental error in the measurements of $v_{g}$ and $v_{s}$.

To conclude, we have studied the steady state of a microtubule assembly in a confined geometry,
where the growth of individual microtubules is restricted in length. We found that, in addition to
the exponentially decaying length distribution in an infinite system, there is a novel steady
state with exponentially increasing distribution of lengths. This prediction is is excellent
agreement with experimental observations in real cells, and is thus a direct verification of
the dynamical instability model of microtubule dynamics.

The authors would like to gratefully acknowledge the Carilion Biomedical Institute for 
providing partial funding for this work. The authors would like
to acknowledge useful discussions with Mitra Feizabadi and Kenneth Meissner.

\end{document}